\documentclass[journal]{IEEEtran}
\usepackage{graphicx}  
\usepackage{subfigure}
\usepackage{multirow}
\usepackage{color}
\usepackage{fancyhdr}
\usepackage{longtable}
\usepackage{parskip}
\usepackage[T1]{fontenc}
\usepackage{dcolumn}   
\usepackage{threeparttable}
\usepackage{bm}        
\usepackage{amsfonts}  
\usepackage{amsmath}   
\usepackage{amssymb}   
\usepackage{lineno}
\ifCLASSINFOpdf
\else
\fi
\begin{document}
%
\title{Curvature-enhanced Graph Convolutional Network for Biomolecular Interaction Prediction}
%
%
%

\author{Cong Shen, Pingjian Ding, Junjie Wee, Jialin Bi, Jiawei Luo and Kelin Xia

\thanks{Manuscript received xxxxxx; revised xxxxxxx.This work was supported in part by the Natural Science Foundation of China (NSFC grant no. 61873089, 62032007), Nanyang Technological University Startup Grant (grant no. M4081842), Singapore Ministry of Education Academic Research fund (grant no. Tier 1 RG109/19, MOE-T2EP20120-0013, MOE-T2EP20220-0010) and China Scholarship Council (CSC grant no.202006130147). \textit{(Corresponding author: Jiawei Luo and Kelin Xia)}}
\thanks{Cong Shen and Jiawei Luo are with the College of Computer Science and Electronic Engineering, Hunan University, Changsha 410000, China (e-mail: cshen@hnu.edu.cn;  luojiawei@hnu.edu.cn)}
\thanks{Pingjian Ding is with the Center for Artificial Intelligence in Drug Discovery, School of Medicine, Case Western Reserve University, Cleveland, OH 44106, USA (e-mail: pxd210@case.edu)}
\thanks{Junjie Wee and Kelin Xia are with the School of Physical and Mathematical Sciences, Nanyang Technological University, 637371, Singapore (e-mail: weej0019@e.ntu.edu.sg; xiakelin@ntu.edu.sg)}
\thanks{Jialin Bi is with the School of Mathematics, Shandong University, Jinan 250100, PR China (e-mail: n2107622c@e.ntu.edu.sg)}}

%
%

\markboth{Journal of \LaTeX\ Class Files,~Vol.~14, No.~8, August~2015}%
{Shell \MakeLowercase{\textit{et al.}}: Bare Demo of IEEEtran.cls for IEEE Journals}
%



\maketitle

\begin{abstract}
Geometric deep learning has demonstrated a great potential in non-Euclidean data analysis. The incorporation of geometric insights into learning architecture is vital to its success. Here we propose a curvature-enhanced graph convolutional network (CGCN) for biomolecular interaction prediction, for the first time. Our CGCN employs Ollivier-Ricci curvature (ORC) to characterize network local structures and to enhance the learning capability of GCNs. More specifically, ORCs are evaluated based on the local topology from node neighborhoods, and further used as weights for the feature aggregation in message-passing procedure. Our CGCN model is extensively validated on fourteen real-world bimolecular interaction networks and a series of simulated data. It has been found that our CGCN can achieve the state-of-the-art results. It outperforms all existing models, as far as we know, in thirteen out of the fourteen real-world datasets and ranks as the second in the rest one. The results from the simulated data show that our CGCN model is superior to the traditional GCN models regardless of the positive-to-negative-curvature ratios, network densities, and network sizes (when larger than 500).
\end{abstract}


%
\IEEEpeerreviewmaketitle

\section{Introduction}
\IEEEPARstart{G}{raphs} or networks are arguably the most commonly used data type for the characterization of topological connections of various objects \cite{defferrard2016convolutional,gilmer2017neural}, ranging from microscale systems, such as molecule structures, cellular structures, atomic/molecular/cellar interactions, functional networks in brain imaging, regulatory networks in genetics, and so on, to macroscale systems, such as social networks, citation networks, transportation networks, etc. Dramatically different from the sequence or image data (Euclidean data), graph data has highly complicated topological structures, which usually directly determine the data functions or properties. Recently, various graph neural network (GNN) models have been proposed to learn the information from the graph data. Based on their tasks, these GNN models can be divided into several types, including node classification \cite{wang2020nodeaug}, link prediction\cite{zhang2019heterogeneous,zhang2020learning}, graph classification \cite{wu2019net,zhang2018end}, and graph property prediction \cite{jiang2021could,li2019deepchemstable,feinberg2018potentialnet}. These GNN models have demonstrated great potential in graph data analysis \cite {wu2020comprehensive,zhang2019graph,zhou2020graph}.

The two essential components of all GNNs are node neighborhood and feature aggregation. For node neighborhood, its most commonly-used definition is that for a certain node (or vertex), all the other nodes that directly connected (through one edge) with this specific node are known as its neighbors. This definition is widely used in GNNs, including Graph Convolutional Network (GCN) \cite{welling2016semi}, Graph Attention Network (GAT) \cite{velivckovic2017graph}, and Graph Isomorphism Network (GIN) \cite{leskovec2019powerful}. The neighborhood of a node can also be defined from random walk methods, in which all the nodes in the path of a random walk are regarded as the neighbors of the initial starting node. This definition is used in GNNs, such as HetGNN \cite{zhang2019heterogeneous}. Feature aggregation, which is key to message passing, is to systematically aggregate the node features (i.e., feature vectors) to update node representations. In general, there are two types of feature aggregation. First, features are aggregated with equal importance. This approach is widely used in models, including GIN \cite{leskovec2019powerful}, GraphSAGE \cite{hamilton2017inductive}, and Neural FPs \cite{duvenaud2015convolutional}. Second, features are aggregated with different weights. In GCN \cite{welling2016semi}, the weights are determined by node degrees. In GAT \cite{velivckovic2017graph}, the weights are evaluated through an attention mechanism, in which feature vectors of the node and its neighbors are multiplied to calculate the weight (or importance) of the neighboring nodes to the specific node.

\begin{figure}[htp]\label{fig:ORC}
\includegraphics[width=3.4in]{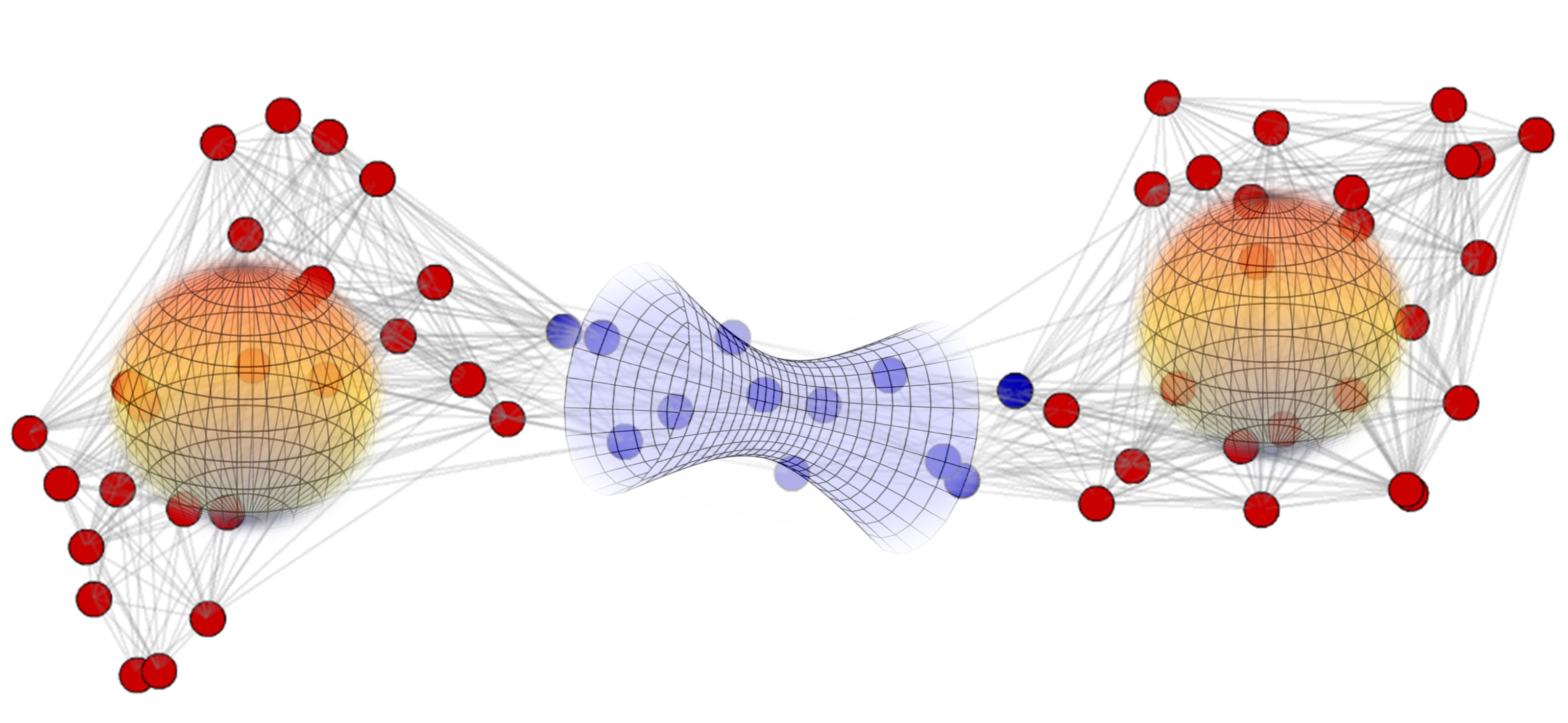}
\caption{\label{setup}
\textbf{Illustration of Ollivier-Ricci curvature (ORC) and "over-squashing" phenomenon.} Here the node ORCs (average of edge ORCs) are plotted. The red and blue colors represent positive and negative ORCs respectively. It can be seen that negative ORCs are found at bottleneck regions, which are related to  "over-squashing" phenomenon \cite{topping2021understanding} }\label{fig:Ollivier-Ricci curvatures}
\end{figure}

Geometric deep learning models have been proposed to incorporate geometric information into deep learning architectures \cite{bronstein2017geometric,bronstein2021geometric,atz2021geometric}. As one of the fundamental concepts in differential geometry, Ricci curvature characterizes the intrinsical properties of manifold surfaces \cite{jost2008riemannian,najman2017modern}. Ricci curvature measures growth of volumes of distance balls, transportation distances between balls, divergence of geodesics, and meeting probabilities of coupled random walks \cite{samal2018comparative}. For two dimensional manifold, Ricci curvature reduces to the classical Gauss curvature. Ricci curvature-based Ricci flow model is key to the proof of Poincar\'{e} conjecture \cite{perelman2003ricci}. Recently, discrete Ricci curvature forms, including Ollivier-Ricci curvature (ORC) \cite{bakry1985diffusions,chung1996logarithmic,sturm2006geometry,ollivier2007ricci,lott2009ricci,ollivier2009ricci,bonciocat2009mass} and Forman Ricci curvature (FRC) \cite{forman2003bochner,sreejith2016forman,samal2018comparative,saucan2018forman}, have been developed and widely used in applications, such as internet topology \cite{ni2015ricci}, community detection \cite{ni2019community,sia2019ollivier}, market fragility and systemic risk \cite{sandhu2016ricci}, cancer networks \cite{sandhu2015graph}, brain structural connectivity \cite{farooq2019network}, and biomolecular systems \cite{wee2021forman,wee2021ollivier}. In particular, discrete Ricci curvatures have been used in the characterization of "over-squashing" phenomenon \cite{topping2021understanding}, which happens at the bottleneck region of a network when the messages propagated from distant nodes distort significantly. More specifically, over-squashing effects emerge from the message-passing-based GNNs if the learned task requires long-range dependencies, i.e., the task depends on representations of distant nodes interacting with each other, and graph has bottleneck regions which result in exponentially many long-range neighboring nodes \cite{topping2021understanding}. Ricci curvatures with negative values are found to be responsible for over-squashing \cite{topping2021understanding}. Figure \ref{fig:ORC} illustrates Ollivier-Ricci curvature (for nodes) and the "over-squashing" phenomenon. More recently, curvature-based graph neural network models have been developed by the incorporation of ORCs into GNN models \cite{ye2019curvature,li2022curvature}. These models have achieved great success in various synthetic and real-world graphs, from social networks, coauthor networks, citation networks, and Amazon co-purchase graph. The curvature graph network model can significantly outperform state-of-the-art(SOTA) when the underlying graphs are of large-sized and dense.

Here we propose a Curvature-enhanced Graph Convolutional Networks (CGCN) for biomolecular interaction prediction. In our CGCN model, the ORC is calculated for each edge of the molecular interaction graph. A ORC related function is used as a weight in the node feature aggregation. In this way, the "geometric information", in particular, for the ``over-squashing'' areas, can be naturally incorporated into our CGCN model. Our model has been systematically compared with eight SOTA models on fourteen commonly used molecular interaction datasets. It has been found that the proposed model can outperform all SOTA models. Further simulation tests are employed to explore the applicability of our CGCN model. It has been found that the CGCN model consistently delivers better results than traditional GCN model. This performance is highly robust to both network densities and ratios between positive ORCs and negative ORCs. Further, consistently with previous results, our ORC-based CGCN model has a better performance on molecular interaction graphs of medium or large sizes (i.e., >500 nodes). Our CGCN does not have obvious advantage and may even have inferior results for graphs with small sizes (<500 nodes).

\section{Related Works}

\subsection{Ollivier-Ricci curvature for graph data analysis}

Ollivier-Ricci curvature is a discrete Ricci curvature model that is developed for the analysis of non-Euclidean data, in particular, graph data. ORC has been used in various applications, such as internet topology \cite{ni2015ricci}, community detection \cite{ni2019community,sia2019ollivier}, market fragility and systemic risk \cite{sandhu2016ricci}, cancer networks \cite{sandhu2015graph}, and brain structural connectivity \cite{farooq2019network}. It has been combined with deep learning models and demonstrated great advantages. A major reason is that Ricci curvature is found to be related to "over-squashing" phenomenon in message aggregation process and can be used to alleviate information distort in message-passing-based GNNs\cite{topping2021understanding}. RicciNet has been developed to identify the salient computational paths with Ricci curvature-guided pruning \cite{glass2020riccinets}. A Ricci flow process, which is parameterized by a reinforcement learning controller, is employed to deform the discrete space of the graph by the systematical removing of edges with negative Ricci curvatures. Curvature Graph Network (CurvGN) has been proposed to incorporate the Ricci curvature information into graph convolutional network so that it can adapt to different local structural topology \cite{ye2019curvature}. An ORC-based message-passing operator is developed by the aggregation of node representations with an ORC-related weight factor, which is obtained through a multi-layer perceptron (MLP) with ORC as its input. Further, Curvature Graph Neural Network (CGNN) has been developed to increase topological adaptivity of GNNs \cite{li2022curvature}. Similar to CurvGN, ORC information is transformed into the weights. However, negative curvature transformation module and curvature normalization nodule are used, so that the relative magnitude of curvature, i.e., large/small curvatures correspond to large/small weights, is well preserved.

Curvature has also been employed in the characterization of embedding spaces. Curvature Graph Generative Adversarial Network (CurvGAN) has been proposed to better preserve the topological properties and alleviate topological distortions \cite{li2022curvature2}. In CurvGAN, global topology of the graph data is approximated by a Riemannian geometric space with constant curvature and local heterogeneous topology is characterized by ORCs. Hyperbolic Curvature Graph Neural Network (HCGNN) integrates discrete and continuous curvature together to enhance hyperbolic geometric learning \cite{yang2022hyperbolic}. Similar to CurvGAN, global topology is characterized by constant curvature manifold and local heterogeneous topology by ORCs. However, in HCGNN, the embedding space is modeled by a hyperbolic space with constant curvature, and ORCs is incorporated into message passing operator though hyperbolic curvature-aware message propagation and ORC-based homophily constraint. Other discrete curvature models have also been employed in learning models, including curvature-informed multi-task learning for graph networks \cite{new2022curvatureinformed}, mixed-curvature multi-relational graph neural network for knowledge graph completion \cite{wang2021mixed}, adaptive curvature exploration hyperbolic graph neural network (ACE-HGNN) \cite{fu2021ace}, etc.

\subsection{Graph Neural Network for Molecular Interaction Prediction}
%
Recently, the application of graph neural networks in multifarious molecular interaction prediction tasks has received increasing attention. For instance, SkipGNN \cite{huang2020skipgnn} utilizes a skip graph neural network to predict molecular interactions. MR-GNN \cite{xu2019mr} infers the interaction between two entities via a dual graph neural network. CSGNN \cite{zhao2021csgnn} uses a contrastive self-supervised graph neural network to predict molecular interactions. Besides, some graph neural network models are applied on some specific molecular interactions. KGNN \cite{lin2020kgnn} is a knowledge graph neural network and MIRACLE \cite{wang2021multi} is a multi-view graph contrastive representation learning model, both used to predict drug-drug interactions. KGE\_NFM \cite{ye2021unified} a unified framework for drug-target interaction prediction by combining knowledge graph and recommendation system. IDDkin \cite{shen2020iddkin} is a network-based influence deep diffusion model for kinase inhibitors prediction. InteractionGraphNet \cite{jiang2021interactiongraphnet} is a novel deep graph representation learning framework for accurate protein-ligand interaction predictions.

\section{Method}
This section is devoted to the introduction of Ollivier-Ricci curvature and ORC-based GNN model.

\begin{figure*}[htp]
\centering
\includegraphics[width=6.5in]{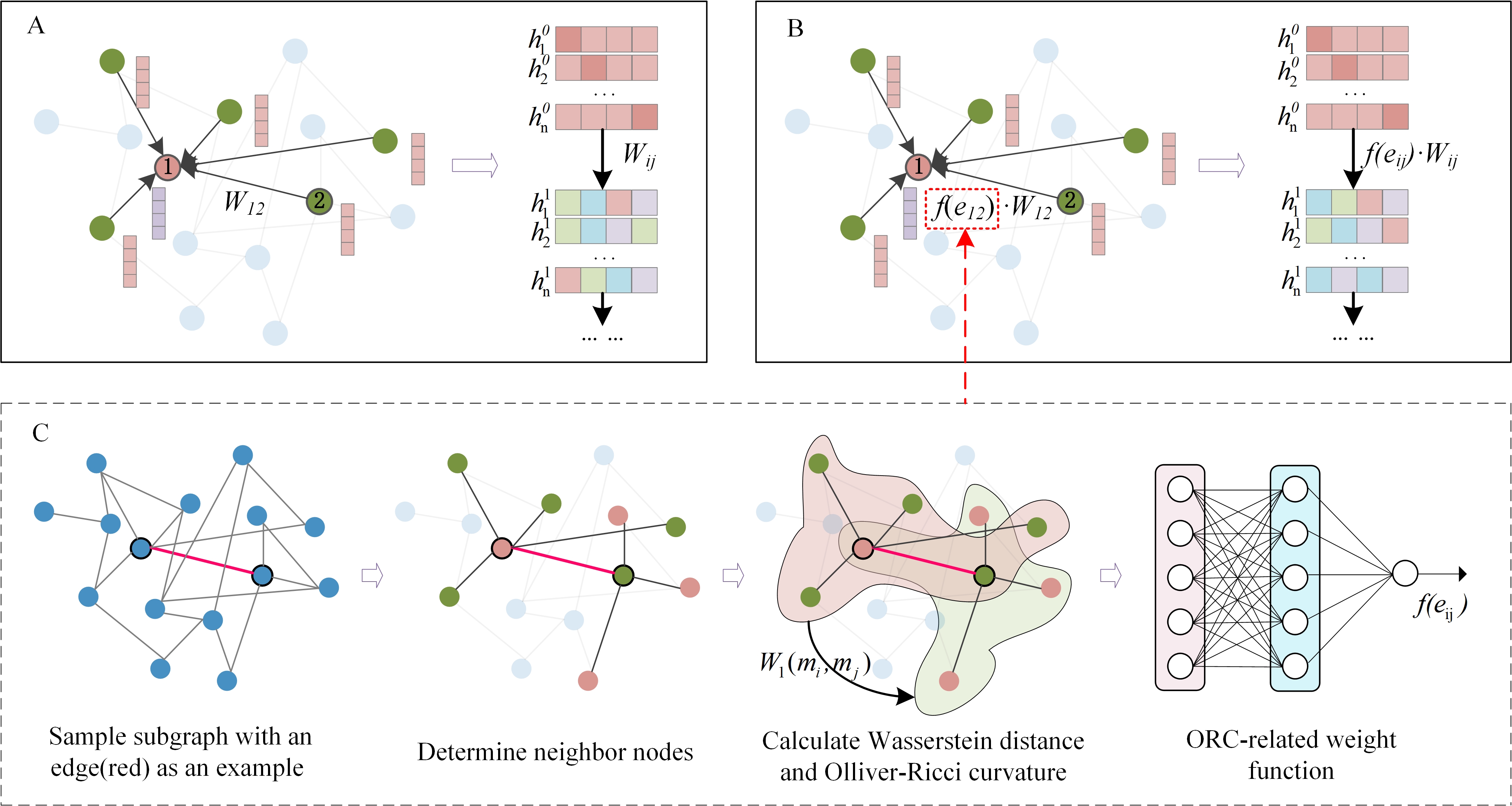}
\caption{\label{setup}
\textbf{An illustration of the ORC-based feature aggregation in message-passing procedure.} \textbf{A} A normal message-passing procedure, in which node feature representation is updated by using all feature vectors from its neighboring nodes with the same weight. \textbf{B} ORC-based message-passing procedure in our CGCN model. The neighboring feature vectors are aggregated with ORC-related weights. \textbf{C} The calculation of ORCs and ORC-related weights. The edge ORC is calculated by using the Wasserstein distance between two probability distributions defined on neighboring nodes. The ORC-related weight is calculated through an MLP. }\label{fig:overview}
\end{figure*}

\subsection{Mathematical Notations}
Here uppercase letters are reserved for matrices (e.g. $W \in \mathbb{R}^{m \times n}$) and lowercase letters are used to denote vectors (e.g. $h \in \mathbb{R}^{d}$).   An interaction network is represented by an undirected graph $\mathcal{G} = \left\{ {\mathcal{V},\mathcal{E}} \right\}$, where $\mathcal{V}$ is the set of vertices and $\mathcal{E}$ is the set of edges. Here $v_{i} \in \mathcal{V}$ is ${i}$-th node and $e_{ij} \in \mathcal{E}$ is the edge between ${i}$th node and ${j}$th node. The edge is formed only when there exists a certain interaction between the two nodes. Further, $c\left({x,y} \right)$ represents the Ollivier-Ricci curvature on edge $e_{xy}$.

\subsection{Ollivier-Ricci Curvature}
Ricci curvature measures the growth of volumes of distance balls, transportation distances between balls, divergence of geodesics, and meeting probabilities of coupled random walks\cite{samal2018comparative}. Ricci curvature equals to the classical Gauss curvature on two dimensional manifold.  Two discrete Ricci curvature forms, i.e., Ollivier Ricci curvature (ORC) \cite{ollivier2007ricci,lott2009ricci,ollivier2009ricci,bonciocat2009mass} and Forman Ricci curvature (FRC) \cite{forman2003bochner,sreejith2016forman}, have been developed. Among them, the most widely used one is ORC, which was originally proposed on metric spaces \cite{ollivier2007ricci,ollivier2009ricci} and further applied to graphs \cite{lin2011ricci,lin2010ricci}. ORC is defined on graph edges. It measures the difference between the edge "distance" (or length of edge) and transportation distance of two probability distributions, which are defined respectively on the two neighborhoods from the-edge-related two vertices. Roughly speaking, positive edge ORC means that there are strong connections (or short "distance") between the two respective neighborhoods, and negative edge ORC indicates weak connections (or long "distance"). It has been found that ORC is also related to various graph invariants, ranging from local measures, such as node degree and clustering coefficient, to global measures, such as betweenness centrality and network connectivity \cite{ni2015ricci}.

Mathematically, for a node $x$ in a graph
$\mathcal{G} = \left\{ {\mathcal{V},\mathcal{E}} \right\}$, its neighbors can be expressed as
$\Gamma(x) = \left\{ {x_{1},x_{2},\cdots,x_{k_{x}}} \right\}$, and the total number of neighbors is $k_x$, which is the degree of node $x$. A probability distribution $m_x$ is defined as,
\begin{align} \nonumber
m_{x}\left( x_{i} \right) = \left\{ \begin{matrix}
{\alpha~~~~~~~~~~~~~~~~~{\rm if}~x_{i} = x~~~~} \\
{{(1 - \alpha)/k_{x}}~~~~~~{\rm if}~x_{i} \in \Gamma(x)} \\
{0~~~~~~~~~~~~~~~~~~{\rm otherwise}~~}, \\
\end{matrix} \right.
\end{align}
where parameter $\alpha \in \lbrack 0,1\rbrack$. Here we use $\alpha = 0.5$, which is the most commonly used value \cite{lin2011ricci,lin2010ricci}. If there is an edge $e_{xy}$ between node $x$ and $y$, a measure $\xi \in \prod\left( m_{x},m_{y} \right)$ between two probability distributions $m_x$ and $m_y$ defines a transportation plan from $m_x$ to $m_y$. This measure is mass-preserving, i.e., $\sum_{y_{j} \in V}{\xi\left( {x,y_{j}} \right) = m_{x}}$ and $\sum_{x_{i} \in V}{\xi\left( {x_{i},y} \right) = m_{y}}$. The amount of mass moved from $x_i$ to $y_j$ is $\xi\left( x_{i},y_{j} \right)$. The $L_1$ Wasserstein distance between $m_x$ and $m_y$, which is minimum average traveling distance and represented by
$W_{1}\left( {m_{x},m_{y}} \right)$, can be computed,
\begin{align}  \nonumber
W_{1}\left( {m_{x},m_{y}} \right) = {\inf\limits_{\xi}{\sum\limits_{x_{i} \in V}{\sum\limits_{y_{j} \in V}{d\left( {x_{i},y_{j}} \right)}}}}\xi\left( x_{i},y_{j} \right).
\end{align}
The Olliver-Ricci curvature on edge $e_{xy}$, denoted as $c\left( {x,y} \right)$, is defined as follows,
\begin{align}  \nonumber
c\left( {x,y} \right) = 1 - \frac{W_{1}\left( {m_{x},m_{y}} \right)}{d(x,y)},
\end{align}
where $d(x,y)$ is the distance between node $x$ and node $y$.

Linear programming (LP) is utilized to calculate Wasserstein distance. Let $\rho\left( x_{i},y_{j} \right) \geq 0$ represent the fraction of "mass" transported for node $x_i$ to $y_I$, the LP formulation can be expressed as follows,
\begin{gather}  \nonumber
   \min{\sum\limits_{x_{i} \in V}{{\sum\limits_{y_{j} \in V}{d\left( {x_{i},y_{j}} \right)\rho\left( {x_{i},y_{j}} \right)}}m_{x}\left( x_{i} \right)}}\\ s.t.:{\sum\limits_{y_{j} \in V}{\rho\left( {x_{i},y_{j}} \right) = 1}},~0 \leq \rho\left( {x_{i},y_{j}} \right) \leq 1 \notag\\
\sum\limits_{x_{i} \in V}{\rho\left( {x_{i},y_{j}} \right)m_{x}\left( x_{i} \right) = m_{y}\left( y_{j} \right)} . \notag
\end{gather}
Note that ORCs are calculated on edges of graphs. The node ORCs are usually defined as the average of edges ORCs. That is for a node $x$, its ORC value is
\begin{align}  \nonumber
c(x) =  \frac{1}{\deg(x)} \sum\limits_{y \in \mathcal{N}{(x)}} c (x,y),
\end{align}
where $\mathcal{N}(x)$ is the neighbors of node $x$, $\deg(x)$ is the node degree of $x$. Figure \ref{fig:ORC} illustrates node ORCs.

\subsection{Curvature Graph Convolutional Network}
In our CGCN model, the ORC information is incorporated into message passing process by using ORC-related edge weights. To alleviate the "over-squashing" effects that usually happen at regions with negative ORC values, we propose an edge weight function that is inversely related to edge ORC values. More specifically, for edge $e_{xy}$ with ORC $c(x,y)$, we define an ORC-related vector,
\begin{equation} \label{eq:ORC_vector}
G_{c}(x,y)=\left( \frac{1 + e^{- c{({x,y})}}}{2},\frac{1 + e^{- 2*c{({x,y})}}}{2},...,\frac{1 + e^{- N*c{({x,y})}}}{2} \right)^T,
\end{equation}
here $N$ is a positive integer values. The edge weight function is defined as follows,
\begin{align}  \nonumber
f_c(x, y) = \tau\left( {W_{\rm MLP}^T G_{c}(x,y)+b} \right)
\end{align}
here $\tau( \cdot )$ represents activate function, $W_{\rm MLP}$ is an weight vector with size $N\times 1$, and $b$ is a constant parameter. Essentially, we use an MLP to learn the edge weight function from the ORC-related vector.

The weight function is then incorporated into the message-passing process, in which node representation is updated by the aggregation of node features from all its neighbors. In our CGCN, the contribution from neighboring node features is not aggregated with equal weight, instead it is scaled by the edge weighted function as follows,
\begin{align}  \nonumber
h_{x}^{l} = \sigma\left( {\sum\limits_{y \in \mathcal{N}{(x)} \cup {\{ x\}}}{\frac{1}{\sqrt{\deg(x)}  \sqrt{\deg(y)}} f_c(x, y) W_{\rm GCN}h_{y}^{l - 1}}} \right)
\end{align}
where $\mathcal{N}(x)$ is the neighbors of node $x$, $deg(x)$ represents the degree of node $x$, $h_{x}^{l}$ and $h_{y}^{l - 1}$ are the node features of $x$ and $y$ after $l$ and $l-1$ message-passing iterations respectively.  Figure \ref{fig:overview} illustrates the ORC-based message-passing process in our CGCN model.

After the message-passing process, node representations $h_{x}$ and $h_{y}$ are obtained for nodes $x$ and $y$. The probability that there is an interaction between two nodes  $x$ and $y$ can be evaluated through a MLP and a hidden layer based score function as follows,
\begin{align}  \nonumber
{\hat{p}}_{xy} = {\rm MLP}\left( \left\| \left( h_{x} + h_{y},{~h}_{x}\odot h_{y}{,~h}_{x}{,~h}_{y} \right) \right. \right),
\end{align}
where $\odot$ is element-wise product, $\left\| (\cdot) \right.$ means the concatenation and the output is a vector, ${\rm MLP}$ is a multi-layer perceptron, and ${\hat{p}}_{xy}$ is the prediction of the relationship between two nodes $x$ and $y$.

\section{Experiments}
This section covers three tests, including a link prediction test on 14 real-world biomedical networks, a simulation test, and a visualization test. Our CGCN model shows good performance on all three tests.
\subsection{Experimental Setup}
\textbf{Datasets.} In this study, two types of graph datasets, i.e., 14 real-world graph datasets and 81 simulated graph datasets, are employed.
The 14 datasets includes ChCh-Miner, ChG-Miner, DCh-Miner, PPT-Ohmnet, DG-AssocMiner,  HuRI-PPI, PP-Decagon, PP-Pathways, CPI\_human, CPI\_celegans, Drugbank\_DTI, Drugbank\_DDI, AdverseDDI, and DisGeNET. These datasets cover various types of biomolecular interaction, including drug-drug interaction networks (ChCh-Miner and Drugbank\_DDI, AdverseDDI), drug-gene interaction networks (ChG-Miner), disease-drug interaction networks (DCh-Miner), protein-protein interaction networks (PPT-Ohmnet, HuRI-PPI, PP-Decagon, and PP-Pathways), disease-gene interaction networks (DG-AssocMiner and DisGeNET), compound-protein interaction network (CPI\_human and CPI\_celegans), and drug-target interaction network (Drugbank\_DTI). Among them, ChCh-Miner, ChG-Miner, DCh-Miner, PPT-Ohmnet, DG-AssocMiner, and HuRI-PPI are obtained from Ref \cite{luck2020reference}. PP-Decagon and PP-Pathways are from Ref \cite{biosnapnets}. CPI\_human and CPI\_celegans are from Ref \cite{tsubaki2019compound}. Drugbank\_DTI, Drugbank\_DDI, and AdverseDDI are from Ref \cite{wishart2018drugbank}. AdverseDDI is taken from Ref \cite{zhu2021multi} and DisGeNET is from Ref \cite{pinero2016disgenet}.

The details of these 14 networks are shown in Table \ref{tab:Statistics of networks}, including the number of nodes and edges, average degree, density, and ratio of positive and negative curvature.

\begin{table}[htb]
\centering
\caption{\textbf{Statistics for 14 biomolecular interaction networks.}}\label{tab:Statistics of networks}
\begin{threeparttable}
\begin{tabular}{llllll}
\hline
Datasets & Nodes & Edges & Degree & Density & Ratio\tnote{1} \\
 \hline
ChCh-Miner\cite{biosnapnets}&	1514&	48514&	64.09&	4.24\%&	1.66\\
ChG-Miner\cite{biosnapnets}&	7341&	15138&	4.12&	0.06\%&	0.61\\
DCh-Miner\cite{biosnapnets}&	7197&	466656&	129.65&	1.80\%&	2.33\\
PPT-Ohmnet\cite{biosnapnets}&	4510&	70338&	31.19&	0.69\%&	0.20\\
DG-AssocMiner\cite{biosnapnets}&	7813&	21357&	5.47&	0.07\%&	0.24\\
PP-Decagon\cite{biosnapnets}&	19081&	715612&	75.01&	0.39\%&	0.50\\
PP-Pathways\cite{biosnapnets}&	21557&	342353&	31.76&	0.15\%&	0.19\\
HuRI-PPI\cite{luck2020reference}&	5604&	23322&	8.32&	0.15\%&	0.18\\
CPI\_human\cite{tsubaki2019compound}&	2013&	2633&	2.62&	0.13\%&	1.21\\
CPI\_celegans\cite{tsubaki2019compound}&	1782&	2659&	2.98&	0.17\%&	 0.94\\
Drugbank\_DTI\cite{wishart2018drugbank}&	12566&	18866&	3.00&	0.02\%&	 0.98\\
Drugbank\_DDI\cite{wishart2018drugbank}&	1977&	563438&	569.99&	28.85\%&	 383.79\\
AdverseDDI\cite{zhu2021multi}&	393&	12480&	63.51&	16.20\%&	87.24\\
DisGeNET\cite{pinero2016disgenet}&	19783&	81746&	8.26&	0.04\%&	0.20\\
\hline
\end{tabular}
\begin{tablenotes}
\footnotesize
\item[1] The ratios between positive ORCs and negative ORCs.
\end{tablenotes}
\end{threeparttable}
\end{table}

\textbf{Baselines.} We compare our CGCN model with 8 state-of-the-art methods, which can be categorized into two classes, i.e., GNN models and network embedding models.
Four GNN models are considered, including Graph Convolutional Network(GCN) \cite{welling2016semi}, Graph Attention Network(GAT) \cite{velivckovic2017graph}, CSGNN \cite{zhao2021csgnn} and SkipGNN \cite{huang2020skipgnn}. Network embedding models are to represent a high-dimensional, sparse vector space with a low-dimensional, dense vector space. They are widely used for network representation learning. Four classical network embedding models are selected, including DeepWalk \cite{perozzi2014deepwalk}, LINE \cite{tang2015line}, Node2Vec \cite{grover2016node2vec} and SDNE \cite{wang2016structural}.

\textbf{Implementation Details.} In this study, we randomly select as many negative samples as there are positive samples, and the whole data set was divided into training set, validation set and test set according to the ratio of 7:1:2. The area under the receiver operating characteristic curve (AUC) and the area under the precision-recall curve (AUPR) are used to evaluate the performance of model. We run each test 10 times, and use the average values as final results. For the simulated network, in order to have a more reasonable topological structure, five disjoint communities are created. The probability of node connection within the community is $p$, and the probability of node connection between communities is $q$. Three types of tests are conducted. First, we fix the node number to be 1000 and systematically change $p$ and $q$ to generate a series of networks with various ratios of positive to negative ORCs. Second, we systematically change the number of nodes from 200 to 20,000 (while keeping $p$ and $q$ to be 0.1 and 0.0001, respectively). Third, we fix the number of nodes to be 1000 and 2000, and systematically change the network density, i.e., the ratio of edge number to the number of all possible edges (in a complete graph).


We use batch size 128 with Adam optimizer of learning rate 5e-4 and run CGCN model in PyTorch. For training, we use a server with 2 Intel(R) Core (TM) I9-10900X 3.70GHz CPUs, 64GB RAM and 2 NVIDIA GeForce RTX 2070 GPUs. For more detailed parameter introduction of the model, please refer to the source code\footnote{A reference implementation of CGCN may be found at  https://github.com/CS-BIO/CGCN}.

\begin{table*}[ht]
\centering
\caption{\textbf{The comparison between CGCN and four GNN methods.}}\label{tab:results of four GNN methods}
\begin{threeparttable}
\begin{tabular}{ lllllllllll }
\hline
\multirow{2}{*}{Datasets}
&\multicolumn{2}{l}{CGCN}&\multicolumn{2}{l}{GCN}&\multicolumn{2}{l}{GAT}&\multicolumn{2}{l}{CSGNN}&\multicolumn{2}{l}{SkipGNN}\\
\cline{2-11}
&AUC&	AUPR&	AUC&	AUPR&	AUC&	AUPR&	AUC&	AUPR&	AUC&	AUPR \\
 \hline
ChCh-Miner&	    \textbf{0.9426}&	\textbf{0.9329}&	0.8984&	0.8791&	0.8786&	 0.8502&	0.9350&	0.9210&	0.8819&	0.8594\\
ChG-Miner&	    \textbf{0.9644}&	\textbf{0.9627}&	0.9352&	0.9409&	0.9514&	 0.9499&	0.9258&	0.9307&	0.9526&	0.9524\\
DCh-Miner&	    \textbf{0.9972}&	\textbf{0.9967}&	0.9966&	0.9961&	0.9966&	 0.9959&	0.9914&	0.9903&	0.8446&	0.8606\\
PPT-Ohmnet&	    \textbf{0.9143}&	\textbf{0.9174}&	0.8937&	0.8988&	0.8798&	 0.8806&	0.9031&	0.9055&	0.8091&	0.7896\\
DG-AssocMiner&	\textbf{0.9945}&	\textbf{0.9925}&	0.9930&	0.9906&	0.9936&	 0.9916&	0.9919&	0.9896&	0.8585&	0.6679\\
PP-Decagon&	    \textbf{0.9397}&	\textbf{0.9402}&	0.9138&	0.9126&	0.8836&	 0.8740&	NA\tnote{1}&NA\tnote{1}&0.8892&	0.8819\\
PP-Pathways&	\textbf{0.9487}&	\textbf{0.9453}&	0.9394&	0.9370&	0.9225&	 0.9177&	NA\tnote{1}&NA\tnote{1}&0.9263&	0.9228\\
HuRI-PPI&	    \textbf{0.9327}&	\textbf{0.9333}&	0.9164&	0.9189&	0.8994&	 0.8965&	0.9228&	0.9269&	0.9119&	0.9182\\
CPI\_human&	    \textbf{0.9738}&	\textbf{0.9770}&	0.9423&	0.9554&	0.9578&	 0.9653&	0.9696&	0.9708&	0.6232&	0.6245\\
CPI\_celegans&	\textbf{0.9886}&	\textbf{0.9891}&	0.9552&	0.9661&	0.9722&	 0.9774&	0.9839&	0.9852&	0.7217&	0.6995\\
Drugbank\_DTI&	\textbf{0.9750}&	\textbf{0.9730}&	0.9234&	0.9371&	0.9476&	 0.9533&	0.9737&	0.9730&	0.8946&	0.6764\\
Drugbank\_DDI&	\textbf{0.9655}&	\textbf{0.9678}&	0.9009&	0.8949&	0.9448&	 0.9514&	0.9537&	0.9495&	0.8144&	0.7772\\
AdverseDDI&	    0.9466&	0.9411&	0.9492&	0.9445&	0.9381&	0.9325&	\textbf{0.9540}&	 \textbf{0.9508}&	0.8450&	0.7610\\
DisGeNET&	\textbf{0.9895}&	\textbf{0.9901}&	0.9723&	0.9785&	0.9829&	 0.9849&	0.9869&	0.9880&	0.9145&	0.9271\\

\hline
\end{tabular}
\begin{tablenotes}
\footnotesize
\item[1] NA indicates that the CSGCN model requires too much memory on the PP-Decagon and PP-Pathways datasets.
\end{tablenotes}
\end{threeparttable}
\end{table*}

\begin{table*}[ht]
\centering
\caption{\textbf{The comparison between CGCN and four network embedding methods.}}\label{tab:results of four network embedding methods}
\begin{tabular}{ lllllllllll }
\hline
\multirow{2}{*}{Datasets}
&\multicolumn{2}{l}{CGCN}&\multicolumn{2}{l}{SDNE}&\multicolumn{2}{l}{Node2Vec}&\multicolumn{2}{l}{LINE}&\multicolumn{2}{l}{DeepWalk}\\
\cline{2-11}
&AUC&	AUPR&	AUC&	AUPR&	AUC&	AUPR&	AUC&	AUPR&	AUC&	AUPR \\
 \hline
ChCh-Miner&	    \textbf{0.9426}&	\textbf{0.9329}&	0.8560&	0.8375&	0.8668&	 0.8431&	0.8436&	0.8424&	0.6881&	0.6736\\
ChG-Miner&	    \textbf{0.9644}&	\textbf{0.9627}&	0.6108&	0.6114&	0.9144&	 0.8943&	0.7312&	0.7354&	0.7623&	0.8159\\
DCh-Miner&	    \textbf{0.9972}&	\textbf{0.9967}&	0.7769&	0.7889&	0.8020&	 0.8077&	0.7494&	0.7461&	0.6279&	0.6243\\
PPT-Ohmnet&	    \textbf{0.9143}&	\textbf{0.9174}&	0.8652&	0.8694&	0.7608&	 0.7675&	0.7118&	0.7411&	0.6274&	0.6409\\
DG-AssocMiner&	\textbf{0.9945}&	\textbf{0.9925}&	0.5831&	0.5797&	0.8461&	 0.8272&	0.6277&	0.6323&	0.7011&	0.7201\\
HuRI-PPI&	    \textbf{0.9327}&	\textbf{0.9333}&	0.9243&	0.9324&	0.8286&	 0.8300&	0.7179&	0.7400&	0.6707&	0.6866\\
PP-Decagon&	    \textbf{0.9397}&	\textbf{0.9402}&	0.8812&	0.8810&	0.8309&	 0.8306&	0.8159&	0.8258&	0.6279&	0.6216\\
PP-Pathways&	\textbf{0.9487}&	\textbf{0.9453}&	0.9115&	0.9116&	0.7678&	 0.7786&	0.8253&	0.8283&	0.6300&	0.6280\\
CPI\_human&	    \textbf{0.9738}&	\textbf{0.9770}&	0.9613&	0.9714&	0.9523&	 0.9441&	0.7600&	0.7798&	0.8620&	0.8876\\
CPI\_celegans&	\textbf{0.9886}&	\textbf{0.9891}&	0.9793&	0.9826&	0.9706&	 0.9697&	0.8331&	0.8558&	0.8388&	0.8303\\
Drugbank\_DTI&	\textbf{0.9750}&	\textbf{0.9730}&	0.7109&	0.6988&	0.9634&	 0.9481&	0.5725&	0.6037&	0.8351&	0.8819\\
Drugbank\_DDI&	\textbf{0.9655}&	\textbf{0.9678}&	0.8048&	0.7776&	0.8085&	 0.7804&	0.7785&	0.7523&	0.7265&	0.6926\\
AdverseDDI&	    \textbf{0.9466}&	\textbf{0.9411}&	0.8945&	0.8630&	0.8954&	 0.8810&	0.8758&	0.8482&	0.7829&	0.7460\\
DisGeNET&	\textbf{0.9895}&	\textbf{0.9901}&	0.6831&	0.6520&	0.8821&	 0.8725&	0.6801&	0.6688&	0.6995&	0.7210\\
\hline
\end{tabular}
\end{table*}

\subsection{Performance on real biomedical datasets}
In this section, we conduct tests to compare CGCN with all the baselines on 14 real-world biomedical datasets. The AUC and AUPR of various methods in link prediction tasks are shown in Table \ref{tab:results of four GNN methods} and Table \ref{tab:results of four network embedding methods}. Table \ref{tab:results of four GNN methods} lists the results of graph neural network methods and Table \ref{tab:results of four network embedding methods} lists the results of network embedding methods. Our CGCN model performs well on most datasets, achieving the best predictive performance on 13 of 14 datasets. Although these 14 datasets differ greatly in terms of network size, average node degree, density, and ratio of positive to negative curvature, our CGCN shows a consistent good performance. In particular, our CGCN model demonstrates great superiority on datesets of ChCh-Miner, ChG-Miner, Drugbank\_DTI and Drugbank\_DDI, in which the results of CGCN are better than GCN model by 4.9\%, 3.1\%, 4.8\% and 7.2\% respectively. Although on AdverseDDI dataset, the performance of our CGCN is not the best, it is only inferior to CSGCN and better than all other models. In general, CGCN shows strong superiority in comparison with both graph neural network models and network embedding methods. 

\begin{figure*}[htp]
\centering
\includegraphics[width=6.0in]{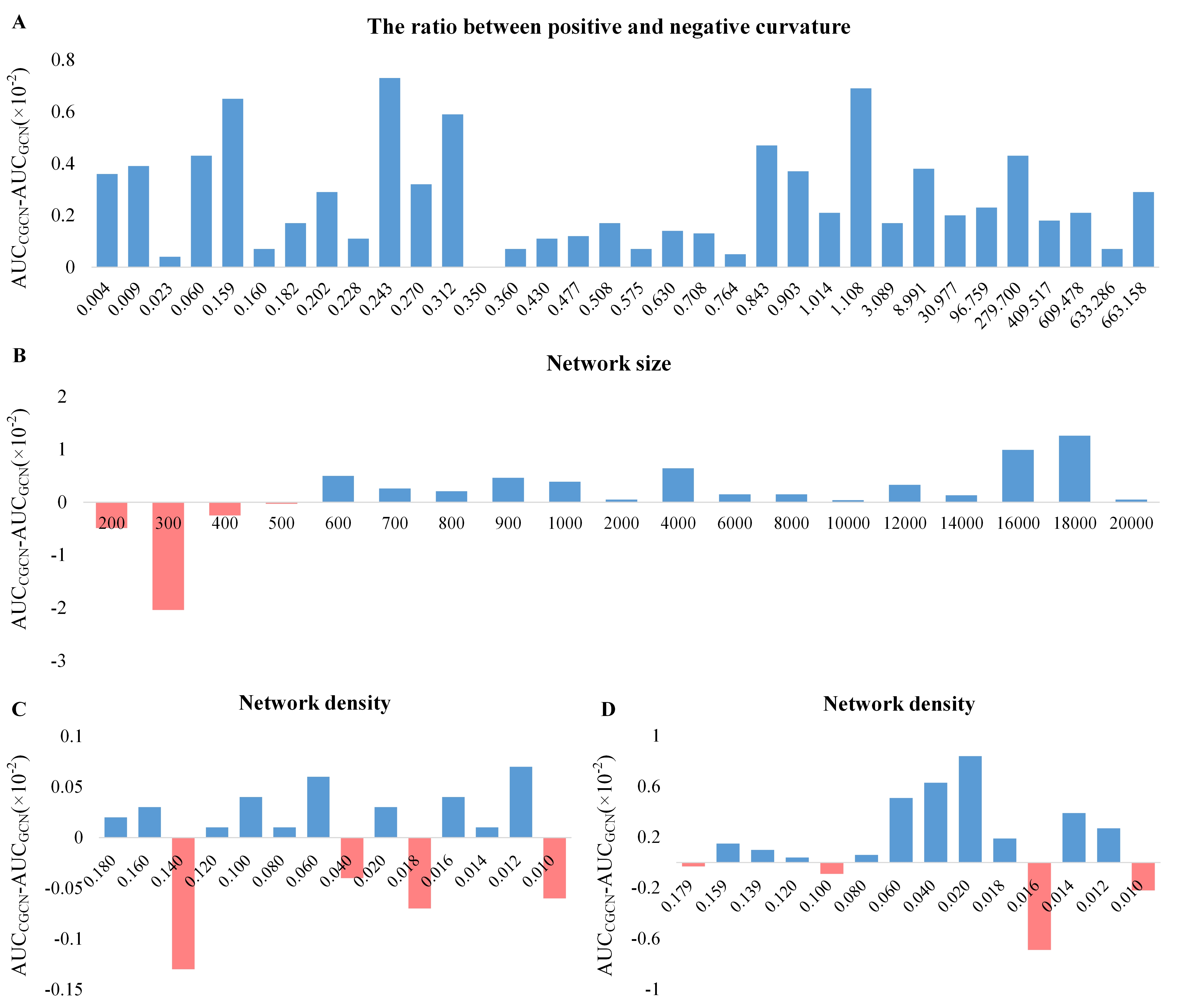}
\caption{\label{setup}
\textbf{The comparison of the results from CGCN and traditional GCN on simulated datasets. } \textbf{A} Performance comparison between CGCN and GCN on different ratios of positive to negative curvatures. \textbf{B} Performance comparison between CGCN and GCN on different network sizes. \textbf{C} Performance comparison between CGCN and GCN on different network densities, when the number of nodes is 1000. \textbf{D} Performance comparison between CGCN and GCN on different network densities, when the number of nodes is 2000. } \label{fig:simulation test}
\end{figure*}

\subsection{Performance on simulated datasets}
In order to further verify the performance of CGCN under various datasets and analyze the limitations of CGCN, we design multiple test cases based on three types of graph properties, including the ratio between positive and negative curvature, network size, and network density. The networks are generated by using the probability of node
connection within the community ($p$) and the probability of node connection between communities ($q$). AUC is used as the metric for the evaluation of the performance. We systematically compare our CGCN model with GCN model \cite{welling2016semi}. The difference between the AUCs in three types of graph property tests are calculated and the results are illustrated in Figure \ref{fig:simulation test}. Note that y-axis represents the difference between the AUCs of CGCN and GCN in all four subfigures, i.e., ${\rm AUC_{CGCN}-AUC_{GCN}}$.


First, we analyze the effect of ratios between positive and negative curvature. We generated 34 random networks with a relatively continuous distribution of positive-to-negative-curvature ratios ranging from 0.004 to more than 600. The results are shown in Figure \ref{fig:simulation test}A. It can be seen that no matter what the positive-to-negative ratios are, our CGCN performance is always better than that of GCN model, which shows that the performance of CGCN is robust to positive-to-negative ratios. In particular, graphs with small positive-to-negative ratios usually have a network topology close to a tree, while large positive-to-negative ratios are associated with complete graphs. The better performance of our CGCN indicates that it is suitable for all kinds of network topologies.

Second, we explore the impact of network sizes on model performance. We set the number of nodes in the simulated network to increase from 200 to 20,000 sequentially. The performance of CGCN and GCN models is shown in Figure \ref{fig:simulation test}B. It can be seen that when the number of nodes in the network is greater than 500, the performance of CGCN is better than that of GCN. When the number of nodes is less than 500, CGCN cannot show obvious superiority, as indicated by the red bars. This indicates that our CGCN model is more suitable for large-sized networks, i.e., nodes larger than 500. This results are consistent with the ones from 14 real-world biomolecular datasets. In fact, our CGCN model is only inferior to traditional GCN model on the AdverseDDI dataset, whose number of nodes is 393 (<500).

Third, we analyze the impact of network densities on model performance. Figure \ref{fig:simulation test}C and \ref{fig:simulation test}D show the prediction performance of the CGCN model under different network densities when the number of network nodes is 1000 and 2000, respectively. It can be seen that the AUC value of the CGCN model is larger than that of the GCN in most networks, indicating that our CGCN model outperforms the traditional GCN model regardless of network densities. This is constant with the observations in Tables  \ref{tab:results of four GNN methods} and \ref{tab:results of four network embedding methods}.

In general, our CGCN model is superior to the traditional GCN model regardless of the positive-to-negative-curvature ratios, network densities, and network sizes (when larger than 500).

\begin{figure*}[htp]
\centering
\includegraphics[width=6.5in]{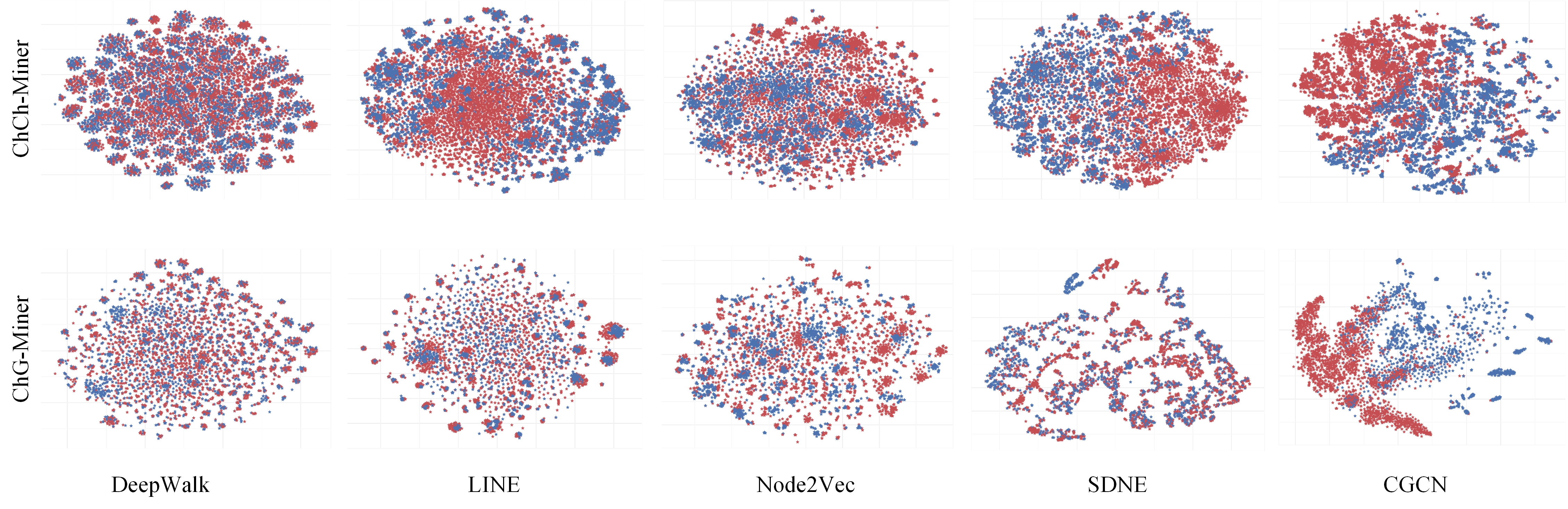}
\caption{\label{setup}
\textbf{The performance of representation learning on ChCh-Miner and ChG-Miner network datasets.} The representation vectors of each nodes in the test datasets are projected into 2D spaces by t-SNE. The red and blue points represent node pairs without link relationship and node pairs with link relationship, respectively. Four network embedding methods are considered in our comparison. }\label{fig:Visualization}
\end{figure*}

\begin{figure*}[htp]
\centering
\includegraphics[width=6.5in]{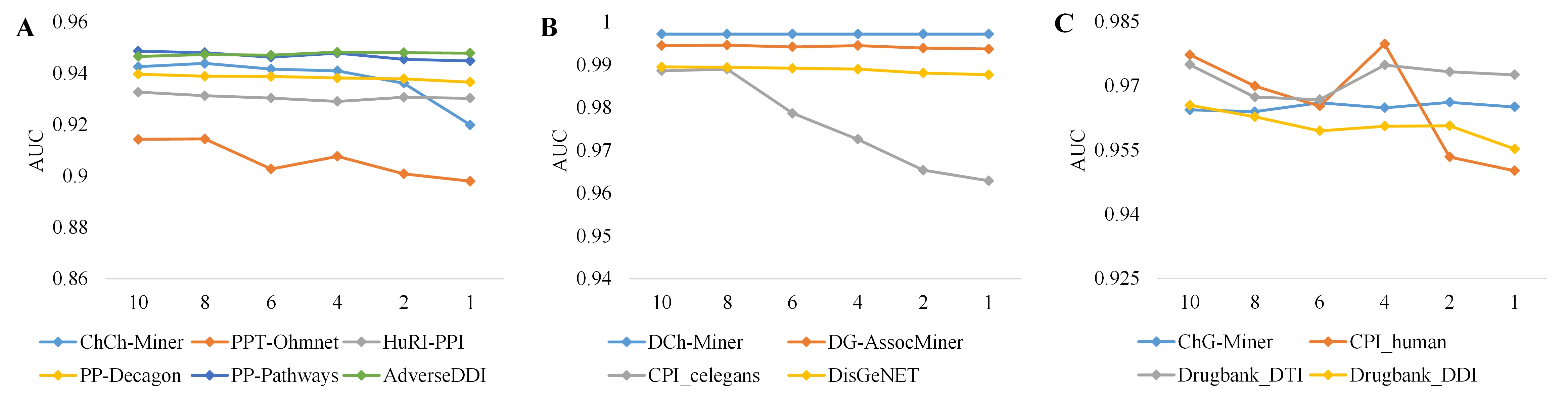}
\caption{\label{setup}
\textbf{Effect of hyperparameter $N$.} \textbf{A} Effect of hyperparameter $N$ on six networks: ChCh-Miner, PPT-Ohmnet, HuRI-PPI, PP-Decagon, PP-Pathways and AdverseDDI. \textbf{B} Effect of hyperparameter $N$ on four networks: DCh-Miner, DG-AssocMiner, CPI\_celegans and DisGeNET. \textbf{C} Effect of hyperparameter $N$ on four networks: ChG-Miner,	 CPI\_human,	Drugbank\_DTI and Drugbank\_DDI.}\label{fig:hyperparameter}
\end{figure*}

\subsection{Performance on representation learning}
In this section, we explore the capabilities of the CGCN model in terms of representation learning. We extract the representation vectors of each nodes in the test datasets and use t-SNE \cite{van2008visualizing} to project the high-dimensional representation vectors into 2D space. The two datasets of ChCh-Miner and ChG-Miner are considered. Our CGCN model is compared with  four network embedding models (DeepWalk, LINE, Node2Vec and SDNE), which usually are used to network representation learning in various tasks. The results are shown in Figure \ref{fig:Visualization}, in which the red and blue points represent node pairs without link relationship and node pairs with link relationship, respectively. It can be seen from the results that the proposed CGCN model is significantly better than the other four network embedding methods in distinguishing node pairs with links and node pairs without links, which shows that the CGCN model has a good capability in representation learning.

\subsection{Parameter Analysis}
We analyze the performance of CGCN by varying the coefficient $N$ to look deeper into the impact of the ORC-related vector in Eq.(\ref{eq:ORC_vector}). We systematically change $N$ from 10 to 1, and the results from the 14 biomolecular interaction datasets are displayed in Figure \ref{fig:hyperparameter}. It can be seen that the size of the $N$ value has a relative small effect (<5\% for all cases) on the performance of our CGCN model. In fact, the variation of datasets including DCh-Miner, DG-AssocMiner,  HuRI-PPI, PP-Decagon, PP-Pathways, Drugbank\_DDI, AdverseDDI, and DisGeNET, are almost neglectable, i.e. less than 1\%. For ChCh-Miner, PPT-Ohmnet, CPI\_celegans and Drugbank\_DDI, with the decrease of $N$ value, the AUC shows a slight decrease.  For the ChG-Miner, CPI\_human and Drugbank\_DTI datasets, the AUC values fluctuate with the decrease of $N$. However, the overall change is not large. 

\section{Conclusion}
The proper incorporation of geometric information into deep learning architectures plays a key role in geometric deep learning models. As one of the fundamental concepts in different geometry, Ricci curvature characterizes the intrinsical properties of manifold surfaces. The discrete Ricci curvatures have found various applications in network and graph data analysis. In particular, they have been used in the characterization of "over-squashing" phenomenon. In this paper, we propose a curvature-enhanced graph convolutional network (CGCN) to incorporate the Ollivier-Ricci curvature (ORC) information into node feature aggregation process. With a better characterization of local topological structures through ORCs, our CGCN model has a more efficient message-passing operator. Experimental results show that the proposed model outperforms the competitive methods in 13 our of 14 real-world biomedical datasets and ranks as second in the rest one. In the simulated tests, our CGCN model is superior to the traditional GCN
model regardless of the positive-to-negative-curvature ratios, network densities, and network sizes (when larger than 500).

\bibliographystyle{IEEEtran}
\bibliography{CGCN}

\end{document}